\documentclass[aps,pre,twocolumn,amsmath,amssymb,superscriptaddress,reprint,longbibliography,nofootinbib]{revtex4-2}

\usepackage[utf8]{inputenc}
\usepackage{graphicx}
\usepackage{color}
\usepackage[colorlinks=true,citecolor=blue]{hyperref}
\usepackage{units}
\usepackage{amsmath}
\usepackage{tabularx}
\usepackage{lipsum}
\usepackage{mathrsfs}
\usepackage{float}

\usepackage{amssymb}
\usepackage{xcolor}
\usepackage{extarrows}
\usepackage{tikz-cd}

\DeclareMathOperator{\sech}{sech}

\newcommand{\bsigma}{\sigma}

\newcommand{\bSigma}{\Sigma}

\AtBeginDocument{%
 \newwrite\bibnotes
 \def\bibnotesext{Notes.bib}
 \immediate\openout\bibnotes=\jobname\bibnotesext
 \immediate\write\bibnotes{@CONTROL{REVTEX41Control}}
 \immediate\write\bibnotes{@CONTROL{%
 apsrev41Control,author="08",editor="1",pages="1",title="0",year="1"}}
 \if@filesw
 \immediate\write\@auxout{\string\citation{apsrev41Control}}%
 \fi
}%

\begin{document}
\title{How Geometrically Frustrated Systems Challenge our Notion of Thermodynamics}
\author{Wolfgang Rudolf Bauer}
\email{bauer\_w@ukw.de}
\affiliation{
Dept. of Internal Medicine I, University Hospital of W\"urzburg,
Oberd\"urrbacher Stra{\ss}e 6,
D-97080 W\"urzburg, Germany, Tel. +49-931-201-39011
}
\altaffiliation[Also at: ]{Comprehensive Heart Failure Center, Am Schwarzenberg 15, A15, D-97080 W\"urzburg, Germany} 

\date{\today}

\begin{abstract}
Although Boltzmann's definition of entropy and temperature are widely accepted, we will show scenarios which apparently are inconsistent with our normal notion of thermodynamics. We focus on generic geometrically frustrated systems (GFSs),  which stay at constant negative Boltzmann temperatures,  independent from their energetic state. Two weakly coupled GFSs at same temperature exhibit, in accordance with energy conservation, the same probability for all energetic combinations.  
Heat flow from a hot GFS to a cooler GFS or an ideal gas increases Boltzmann entropy of the combined system, however the maximum is non-local, which, in contrast to conventional thermodynamics,  implies that both subsystems maintain different temperatures here. 
Re-parametrization can transform these non-local into local maxima with corresponding equivalence of re-defined temperatures. However, these temperatures cannot be assigned solely to a subsystem but describe combinations of both. The non-local maxima of entropy restrict the naive application of the zeroth law of thermodynamics. Reformulated this law is still valid with the consequence that  a GFS at constant negative temperature can measure positive temperatures. Heat exchange between a GFS and a polarized paramagnetic spin gas, i.e. a system which may achieve besides positive also negative temperatures, drives the combined system to a local -, or non-local maximum of entropy, with equivalent or non-equivalent temperatures here. Energetic constraints determine which scenario results.  In case of a local maximum,  the spin gas can measure temperature of the GFS like a usual thermometer, however, this reveals no information about the energetic state of the GFS.  

\begin{description}
\item[PACS numbers] 5.70.Ln, 5.20.Gg
\end{description}            

\end{abstract}
\maketitle

\section{Introduction}
Though thermodynamics is a long-standing discipline within physics, there are still issues which are discussed controversially as the recent dispute whether Boltzmann or Gibbs entropy is correct, showed \cite{buonsante2016dispute}. This controversy is tightly linked to the question about existence or non-existence of negative temperatures \cite{baldovin2021statistical}, as Gibbs entropy provides solely positive ones, whereas Boltzmann's entropy allows negative temperatures, the existence of which was addressed  theoretically and experimentally in early  pioneering works \cite{onsager1949statistical,gauthier2019giant,johnstone2019evolution,ramsey1956thermodynamics,purcell1951nuclear}. 

Gibbs approach reveals consistent thermostatistics, i.e. the equivalence of thermodynamic forces with its counterpart in statistical mechanics  \cite{dunkel2014consistent,hilbert2014thermodynamic}.
The most striking advantage of Boltzmann entropy is that its increase correctly provides the direction of heat flow in combined systems from the hot to the cold subsystem. Defining the relation hot to cold from the Boltzmann temperatures as  $1/T_{hot}<1/T_{cold}$, extends this concept also to negative temperatures \cite{baldovin2021statistical}, which could be proven experimentally already by Abragam and Proctor  \cite{abragam1958,PhysRev.106.160}.
Absence of heat flow, i.e. thermodynamic equilibrium, is present when Boltzmann entropy has reached its maximum and temperatures of the subsystems have become equivalent.  This is the foundation for the zeroth law of thermodynamics, which is the base for thermometry. It states that  if two system are in thermal equilibrium with a third, all are in thermal equilibrium, implying equivalence of temperatures, i.e. temperature is a transitive parameter.    

Though the Boltzmann entropy is in many points superior to Gibbs entropy, its concept may be challenged, and hence, our notion of thermodynamics. In literature the systems which are capable to stay at negative temperature, e.g. spin systems in an external magnetic field, also have the option to reach positive values. When brought in contact with a conventional system, e.g. a gas, both will equilibrate to equal positive temperatures. The question arises what happens, if a subsystem restricted to stay at negative temperatures comes in contact with a subsystem exhibiting solely positive temperatures? Or more generally, if the temperature range the subsystems are restricted to, do not match? How must one reconsider the concept of equal temperatures of subsystems at maximum entropy of the combined system, i.e. the equilibrium state in classical thermodynamics?  What are the implications for the zeroth law of thermodynamics, and, hence thermometry? A perhaps provocative question is, whether a subsystem, the temperature range of which does not match that of the other, may still serve as the latter's thermometer?  

To tackle the above questions we consider as a paradigm generic geometrically frustrated systems, namely $N$ antiferromagnetically interacting spins, which are located at the corners of a regular  $(N-1)-$dimensional  simplex. They have the property that degeneracy of its energy levels decreases monotonically with increasing energy, i.e. temperature is always negative. 

The paper is structured as follows:  first,  the formal and statistical framework is presented in section~
(\ref{section:Thermodynamics}). Here, it becomes explicit that the gradient of the  Boltzmann temperature in composite systems  emerges as a stochastic force,  which drives it to its most probable state, which is not necessarily a local maximum of entropy. The above mentioned GFSs are introduced in section~(\ref{section:SingleGFS}).
With these tools in hand     
the weak coupling of a GFS to another GFS at same and at different temperature is investigated in section~(\ref{section:CoupledGFS}), as well as  in section~(\ref{section:GFS-Gas}) the coupling to an ideal gas, i.e. a system restricted to stay a positive temperatures. Whether a polarized  paramagnetic spin gas, i.e. a system capable to stay at positive and negative temperatures, may measure the temperature of a GFS  is investigated in  section~(\ref{section:GFSPSGas}). In section~(\ref{section:Discussion}) the results are discussed in the framework of the present notions of entropy and temperature. In particular we present strategies for thermometry of systems, which maintain different temperatures under equilibrium conditions, and, by this, generalize the zeroth law of thermodynamics.

\section{Thermodynamic Driving Forces and Boltzmann Entropy} 
\label{section:Thermodynamics}
We consider an isolated  system, the microscopic state of which is characterized by a state variable $\bsigma$, e.g. the configuration of a spin system. The system undergoes stochastic transitions, described by a trajectory $\bsigma(t)$ in state space $\Omega$, for which, for simplicity, it is assumed that it contains a finite number of states. For the stochastic dynamics holds (see also Appendix \ref{appendix1} for more background information about the assumptions) stationarity, irreducibility, and ergodicity. Furthermore, it is assumed that  Gibbs's postulate of equal a priori probability of microstates in equilibrium holds, i.e. $p^{(eq)}(\bsigma)\equiv1/|\Omega |$. For transitions between microstates micro-reversibility,  i.e. detailed balance holds. Together with equal a priori probability this implies equal transition rates between two microstates. 

We now consider subsets of states $\bSigma\subset\Omega$ and transitions between them, which are defined by all transitions of microstates belonging to respective subsets, i.e.  $\bSigma_1\ni\bsigma_1\leftrightharpoons \bsigma_2\in\bSigma_2 $. 
One can quantify the probabilistic drift dynamics  between two subsets from respective conditional transition probabilities   after time $\Delta t $  by a force 
\begin{equation}
\mathfrak{F}_{\bSigma_2,\bSigma_1}(\Delta t)=\ln\left(\frac{p(\bSigma_2|\bSigma_1;\Delta t)}{p(\bSigma_1|\bSigma_2;\Delta t)}\right)\;.\label{driftforce0}
\end{equation}
This force vanishes if the conditional transition between both assemblies  are equal probable, otherwise it provides the direction and magnitude  of the drift. 
Detailed balance  (see Appendix~\ref{appendix1}) relates probabilistic drift dynamics between two assemblies to its equilibrium probabilities $p^{(eq)}(\bSigma)=\sum_{\bsigma\in\bSigma} p^{(eq)}(\bsigma)=|\bSigma|/|\Omega| $ by   
\begin{equation}
 \underbrace{\frac{p(\bSigma_2|\bSigma_1;\Delta t)} {p(\bSigma_1|\bSigma_2;\Delta t)}}_{\text{probalistic drift dynamics}}=\underbrace{\frac{p^{(eq)}(\bSigma_2)}{p^{(eq)}(\bSigma_1)}=\frac{|\bSigma_2|}{|\bSigma_1|}}_{\text{ equilibrium parameters}}\;.\label{ForceEntropy}
 \end{equation}
Hence, the drift force in Eq.~(\ref{driftforce0}) directly derives from the gradient  of Boltzmann like  entropies $S(\bSigma)=\ln(|\bSigma|)$ as
 \begin{equation}
\mathfrak{F}_{\bSigma_2,\bSigma_1}=S(\bSigma_2)-S(\bSigma_1)\;. \label{driftforce}
\end{equation}     
These entropies are referred to as ``Boltzmann like'', as they are defined by the number of states of a subset of $\Omega$. 

If the system is composed of subsystems, the assemblies $\bSigma$  may be related to macroscopic state observables of the latter,  e.g. volumes, magnetizations, or energies, denoted by $\epsilon=(\epsilon_1\cdots \epsilon_n)^t $, with $\epsilon_i$ as the macroscopic variable of subsystem $i$. It can be shown (see Appendix~\ref{appendix2}), that for transitions between assemblies of neighbored state variables,   $\epsilon '=\epsilon+d\epsilon\leftarrow \epsilon$, Eq.~(\ref{driftforce}) translates   to
 \begin{equation}
d \mathfrak{F}_{\epsilon',\epsilon} =dS=\nabla_\epsilon S(\epsilon)\; d\epsilon|_{\text{constraints}} \;,\label{driftforce2}
\end{equation}     
with $\nabla_\epsilon=(\partial_{\epsilon_1},\cdots,\partial_{\epsilon_n})$. The subscript $|_{\text{constraints}} $ denotes, that constraints as e.g. energy conservation must be respected. 
The last Equation makes explicit, that the drift force derives from a potential, namely the entropy, which makes it a conservative force. This  force drives the system to some macroscopic state  $\epsilon_m$ where entropy is at its maximum.  

This is illustrated for the example of energetically weakly coupled subsystems, with energies $E^{(i)}$ , i.e.  $\epsilon=(E^{(1)},\cdots , E^{(n)})$, and total energy $E=\sum_i E^{(i)}$. Here, the entropy decomposes into entropies of the individual subsystems, $S(E)=\sum_i S^{(i)}(E^{(i)}) $ (see Appendix \ref{appendix3}), and Eq.~(\ref{driftforce2})  reads
\begin{equation}
d\mathfrak{F}_{\epsilon',\epsilon}=\sum_{i=1}^{n}1/T^{(i)} dE^{(i)}|_{\text{energy conservation}}\;, \label{driftforce3}
\end{equation}
with Boltzmann temperature $1/T^{(i)}=\partial_{E^{(i)}} S^{(i)}$ of the respective subsystems, and the constraint that energy is conserved, i.e. $\sum_i dE^{(i)}=0$. Note, that for $n=2$ subsystems, this becomes $dE^{(II)}=-dE^{(I)}=\delta E^{(II,I)}$ and the drift force is proportional to the difference of its inverse temperatures  
\begin{equation}
d\mathfrak{F}_{\epsilon',\epsilon}=(1/T^{(II)}-1/T^{(I)}) \delta E^{(II,I)} \;,\label{driftforce4}
\end{equation}
with $ \delta E^{(II,I)} $ as the amount of energy transfer from system $I$ to $II$. Hence $d\mathfrak{F}_{\epsilon',\epsilon} $ directs heat flow towards the colder subsystem, i.e. that with higher inverse temperature. 
It must be stressed here that for the above derivations no presuppositions about the inner structure of the subsystems were made. In particular this means that Eqs.(\ref{driftforce3},\ref{driftforce4}) are not restricted to extensive subsystems, which fulfill the thermodynamic limit. Instead any  subsystem is conceivable, in particular subsystems which consist of strongly interacting elements as the GFSs.

Two scenarios may occur, when the combined system is driven towards the most probable state $\epsilon_m=(E^{(1)}_m,\hdots,E^{(n)}_m) $.  
If the maximum of entropy is local,  the driving force vanishes here , $d\mathfrak{F}_{\epsilon',\epsilon_m}=0 $, and together with the constraint of constant energy, temperatures of the  subsystems become equivalent,
\begin{equation}
1/T^{(i)}(E^{(i)}_m)\equiv \text{constant} \;.
\end{equation}
However, if $\epsilon_m$ is at the boundary of the macroscopic observable space, one may get the situation that entropy is at its maximum, though the the force is not vanishing, and the equivalence of temperatures does not  hold.   

The first scenario is in accordance with our normal notion, of how subsystems, after having brought in thermal contact, behave. During relaxation the entropic force drives the combined system to the most probable state,  at which the subsystems exhibit equal temperature. Fluctuations around this state occur, but they are normally small, which makes the combined system apparently  stay statically in this most probable state. Detailed balance assures that there is no net energy exchange between the subsystems. In phenomenological thermodynamics this most probable static state is then referred to as the equilibrium. However,  in statistical mechanics this most probable state  $\bSigma_{\epsilon_m}$ must not be confused with  equilibrium. In equilibrium, all microstates of $\Omega$  are equally probable. The most probable state is just a subset of state space, or in terms of entropy, this implies that the entropy of the composite system is larger than the sum of entropies of its subsystems, 
\begin{equation}
\sum_{i}^{n} S^{(i)}(E^{(i)}_m)=S(\bSigma_{\epsilon_m})< S(\Omega)\;.\label{convexity}
\end{equation}
Notice that this is an explicit generalization of the convexity property of the entropy function \cite{puglisi2017temperature, cerino2015consistent}.
In daily life experience, the measure of the most probable state is almost identical with that of the whole state space,  which makes the smaller sign in the last Equation to an equal sign, and justifies the statement of thermodynamics that entropy is an additive extensive parameter. 

In contrast, we will show in the next sections that special GFSs behave different from this notion. 
When weakly coupled to other systems, the most probable state may be a ``small'' subset of state space, and the smaller sign in Eq.~(\ref{convexity}) becomes significant. There even may be no most probable state at all, when identical GFSs are coupled. There are also scenarios which exhibit only a non-local maximum of entropy, which implies that both subsystems will never achieve equal temperatures in equilibrium. Though counterintuitive, this behavior will be consistently described in the above framework which relates the stochastic drift force to Boltzmann entropy and temperature. However, both, the non-dominance of the most probables state -, and the non-equivalence of temperatures in equilibrium force to reconsider the zeroth law of thermodynamics and, hence,  concepts of temperature measurement.

\section{Thermodynamic Properties of a Single Geometrically Frustrated System}
\label{section:SingleGFS}
As a special generic GFS we consider an Ising model with  $N$ spins, which are located at the corners of a regular $(N-1)$-dimensional simplex. This is the generalization of primitive geometrically frustrated systems \cite{moessner2001,moessner2006geometrical} in $2$ (regular triangle) -, or $3$ (tetrahedron) dimensions.  Each spin has two orientations $s_i=\pm1$ and the microscopic state variable is 
\begin{equation}
\bsigma=(s_1,\hdots,s_N) \;. \label{Eq:microstate}
\end{equation}
There is a mutual antiferromagnetic interaction  between all spins and the Hamiltonian takes the form 
\begin{eqnarray}
H(\bsigma)&=&\frac{J}{2}\;\sum_{i,j,\;i\neq j}^{N} s_i s_j\\
&=&\frac{J}{2} \bigg[\bigg(\underbrace{\sum_i^{N} s_i}_{=m}\bigg)^2 -\sum_i s_i^2\bigg]\nonumber\\
&=&\frac{J}{2}\;(m^2-N)\;,\label{Hamilton1}
\end{eqnarray}
with interaction energy $J>0$ and magnetic moment $m$, which we define as the difference of up and down orientated spins. To avoid unnecessary constants the magnitude of the dipole moment of the single spin is comprised in the interaction energy. 
The system is allowed to undergo stochastic transitions between microstates by spin-flip processes which respect energy conservation.  The magnitude of the responsible interactions for these spin flip processes are assumed to be much smaller than the mutual antiferromagnetic interactions, i.e. they do not affect measure parameters of state space. This approach is frequently used e.g. for paramagnetic spins subjected to an external field  \cite{abragam1958} and interacting spins in mean field models of magnetism \cite{mukamel2005}.

As the term $-J/2\; N$ in Eq.~(\ref{Hamilton1}) just determines the offset of energy with no further implications, we will omit it. In addition all energetic values are  normalized to the absolute value of the interaction energy $J$.  Hence  re-scaling the Hamiltonian $H\to H/J+1/2 N $  makes energies of one  GFS to be found between
\begin{equation}
E_{min}=0 \le \underbrace{\frac{1}{2}\;m^2}_{=E} \le E_{max}=\frac{1}{2}\;N^2\;.\label{Eq:rescaledE}
\end{equation}
Elementary combinatorics reveals that the set of states $\bSigma_m$ with magnetic moment  $m$ consists of 
\begin{equation}
|\bSigma_m|=\binom{N}{\tfrac{N+m}{2}}\label{Eq:Binomial}
\end{equation}   
elements, and Boltzmann entropy becomes
\begin{equation}
S(\bSigma_m)=\ln\left(\binom{N}{\tfrac{N+m}{2}}\right)\;. \label{Eq:BEntropygfsm}
\end{equation} 
For large $N$ one may approximate the binomial coefficient by a  Gaussian 
\begin{eqnarray}
|\bSigma_m| &\approx & \frac{1}{\sqrt{2\pi}}\;\frac{2^N}{\sqrt{N/4}}\;e^{-\tfrac{1}{2}\;\tfrac{m^2}{N}}\;,\cr\cr\label{Eq:Gaussian}
&=&\frac{2^{N+1}}{\sqrt{2\pi\;N}}\; e^{-\frac{E}{N}}\;.
\end{eqnarray}    
%\begin{widetext}
As both orientations of the magnetic moment, $\pm m$, contribute to the same energy, the Boltzmann entropy of the states with energy $E$ then derives as
\begin{equation}
S(\bSigma_E)=\ln\left( \frac{2^{N+2}}{\sqrt{2\pi\;N}}\  \right)-\frac{E}{ N}\; \label{Eq:BEntropygfs}
\end{equation}
with Boltzmann temperature 
\begin{equation}
\frac{1}{T}=\partial_E \ln(S(\bSigma_E))=-\frac{1}{N}\;.\label{TemperatureGFS}
\end{equation}
This temperature  is constantly negative,i.e. independent from energy, and scales inversely with the  number of spins  $1/N$. Obviously a GFS does not fulfill the properties we assign to a normal system, for which the thermodynamic limit holds. Neither is energy nor entropy additive nor are they extensive, and temperature is not an intensive quantity. Notice, that the Hamiltonian in Eq.~(\ref{Hamilton1}) must not be confused with the similar looking Curie-Weiss Hamiltonian \cite{campa2009statistical}, which, describing also long range interactions, re-scales the above Hamiltonian by the number of spins $N$, and, by this, makes the Hamiltonian an extensive quantity. 

Despite its non-extensive properties, the GFS may be assigned a Boltzmann entropy by its number of states for a certain energy, and from that a temperature is formally derived. Its physical significance as a component of a stochastic force emerges when the GFS couples weakly to another system, as shown in the previous section. In the following examples are analyzed.

\section{Two Weakly Coupled GFS}
\label{section:CoupledGFS}
We assume that the number of spins $N^{(i)}$ of our GFSs is very large, so that the  energy of the GFS is much smaller than its  upper boundary (Eq.~(\ref{Eq:rescaledE})), 
\begin{equation}
\frac{1}{2}\; ( m^{(i)})^2\ll \frac{1}{2}\;  (N^{(i)})^2 \label{Eq:freeenergyex}\;,
\end{equation}
and  the  systems may mutually exchange their energies without constraint.   
If two of these GFSs are brought in weak thermal contact, one has to distinguish the case of systems at same or different temperature. 
\subsection{ Systems at Same Temperature}
Being at same temperature implies with Eq.~(\ref{TemperatureGFS}) for the subsystems $N^{(I)}=N^{(II)}=N $
With respective energies $E^{(I)},\; E^{(II)}$, state space has
\begin{eqnarray}
|\bSigma_{E^{(I)}, E^{(II)}}|&=&|\bSigma^{(I)} | |\bSigma^{(II)}|\cr
&\sim &\exp\left(-\frac{1}{N}\underbrace{(E^{(I)}+E^{(II)}}_{=E\equiv{\text{constant}}})   \right) \label{2IdGFS1}
\end{eqnarray}  
elements, which, as the related entropy  $S(\bSigma_{E^{(I)}, E^{(II)}}) $, becomes a constant. A flat entropy profile resulting from equivalent  temperatures of the two subsystems makes the driving force of Eq.~(\ref{driftforce4}) vanish. So instead of having a stable balance at a point of maximum entropy, there is an indifferent balance over the whole space of accessible states. 
 
This extreme example nicely shows for two systems in weak thermal contact: having the same Boltzmann temperature does not define a unique state of energy of the two systems. And, being in a certain energy state $E^{(I)}_0,\; E^{(II)}_0$ and having identical temperature does not imply that the system is in thermodynamic equilibrium, in which all states with $E^{(I)}+ E^{(II)}=E $ have the same probability. Overall state space $\Omega$ is the unification of all spaces $\bSigma_{E^{(I)}, E^{(II)}}$ with $E^{(I)}+ E^{(II)}=E $, and as these spaces have all the same number of elements (Eq.~(\ref{2IdGFS1})) one gets 
\begin{eqnarray}
|\Omega|&=&\sum_{\scriptscriptstyle{E^{(I)}+ E^{(II)}=E}} |\bSigma_{E^{(I)}, E^{(II)}}|\cr
&=&|\bSigma_{E^{(I)}_0, E^{(II)}_0}|\;\sum_{\scriptscriptstyle{E^{(I)}+ E^{(II)}=E}} \;1\cr
&\sim &|\bSigma_{E^{(I)}_0, E^{(II)}_0}|\sqrt{2 E}  \;,
\end{eqnarray} 
This number may by far exceed  that of the state in Eq.~(\ref{2IdGFS1}). Therefore, there is no single preferential and dominating state in equilibrium. Instead  fluctuations  make every combination of energies, which respect energy conservation,  appear with the same probability, which is definitely not our accustomed notion of an equilibrium state.  
 
 \subsection{Two GFSs at Different Temperature}
\begin{figure*}
\onecolumngrid
\centering
\includegraphics[width=16cm] {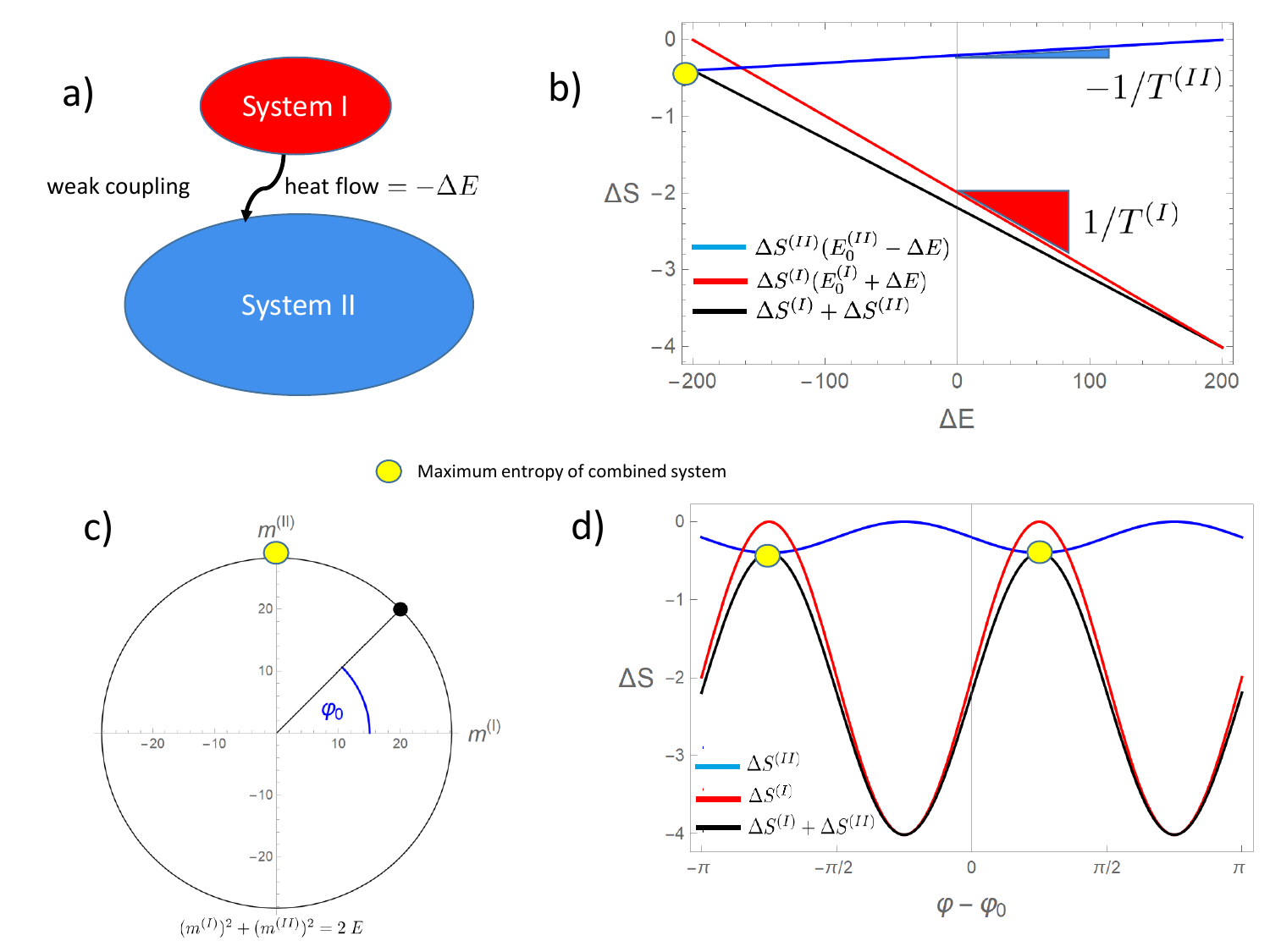}
\caption{Two GFSs in weak thermal contact. a)  Direction of heat flow (Eq.~(\ref{Eq:temperaturegfs})) from the hotter system I ($N^{(I)}=100 $ spins) to system $II$ ( $N^{(II)}=1000 $ spins).  b) Entropies of GFSs and the combined system as a function of heat transfer from system $I $ to $II$, $-\Delta E$. Entropies of the GFSs are normalized to their maximum value, $\Delta S^{(i)}(E^{(i)})=S^{(i)}(E^{(i)})-S^{(i)}(0)$. Initial magnetic moments are $m^{(I)}_0=m^{(II)}_0=20$, and hence initial energies  $ E_0^{(I)}=E_0^{(II)}=E_0=200$.  Maximum entropy, i.e. the most probable state, of the combined system (black line) occurs after delivery of all energy from system $I$, $\Delta E=-E_0=-200 $ (yellow circle) to system $II$. As inverse temperatures are constant throughout (slope of triangles), they differ in particular at this state and slope of  overall entropy does not vanish at its maximum.  c)  Re-parametrization of the state variable  $\Delta E$ by $\varphi=\arctan(m^{(II)}/m^{(I)})$ (Eqs.~(\ref{Eq:mapping00}-\ref{Eq:mapping})). The corresponding space is the compact, unbounded unit circle, $\mathscr{S}^1$, and  $\varphi_0=\arctan(1)=\pi/4$ becomes the new initial state variable. d) In the unbounded $\mathscr{S}^1$ absolute maxima become local ones, i.e. $\partial_\varphi S (\varphi=\pm \pi/2)=0 $. } 
\label{Fig:GFSweakv}
\twocolumngrid
\end{figure*}

We consider two GFSs in weak thermal contact at different temperatures. This is realized by different number of spins $N^{(II)}>N^{(I)}$, i.e. with Eq.~(\ref{TemperatureGFS}) 
\begin{equation}
\underbrace{1/T^{(II)}}_{=-\frac{1}{ N^{(II)}}}>\underbrace{1/T^{(I)}}_{=-\frac{1}{ N^{(I)}}}\;. \label{Eq:temperaturegfs}
\end{equation}
In accordance with our usual notion of thermodynamics, the drift force in Eq.~(\ref{driftforce4}) makes heat flow from the hotter system $I$ to the cooler system $II$ (Fig.~\ref{Fig:GFSweakv}(a, b)). Conservation of energy, and the fact that the number of spins does not limit energy exchange (Eq.~(\ref{Eq:freeenergyex})), define which energies the subsystems may take
\begin{eqnarray}
E^{(I)}&=&E_0^{(I)}+\Delta E \cr
E^{(II)}&=&E_0^{(II)} - \Delta E \cr
&\text{with }&\cr
-E^{(I)}_0 \le & \Delta E& \le E^{(II)}_0 \;,\label{CompactSpace}
\end{eqnarray}
and $ E_0^{(i)}$ as the initial energies of the subsystems. 
The parameter $\Delta E$ becomes a state parameter of the combined system, with the  interval 
\begin{equation}
\mathscr{I}=[-E^{(I)}_0,  E^{(II)}_0] \label{Eq:intervalstatespace}
\end{equation}
 as its compact bounded state space.  
As temperatures of the subsystems remain constant, the resulting drift force in Eq.~(\ref{driftforce4}) is constant as well and directs system I and II, respectively  towards lower and higher energy values (Fig.~\ref{Fig:GFSweakv}(b)),      
\begin{equation}
d\mathfrak{F}_{\scriptscriptstyle{\Delta E+d\Delta E, \Delta E]}}|\mathscr{I}\equiv \text{const}>0,\;\text{if } \Delta E<0 .
\end{equation}    
The whole system is driven  to an absolute maximum of its entropy, located at the boundary of state space  where  system $I$  has delivered all its energy to system $II$
\begin{equation}
\Delta E=-E_0^{(I)}\;,\label{Eq:maxEntropyGFSGFS}
\end{equation}
i.e. in one of its ground states, where its magnetic moment  vanishes
\begin{equation}
m^{(I)}=0\; .\label{Eq:maxEntropyGFSGFS2}
\end{equation}
At this most probable state, the subsystems retain their different temperatures (Fig.~\ref{Fig:GFSweakv} (b)).

Re-parametrization of the state variable $\Delta E$ may transform the absolute, but non-local maximum into a local one.  Energy conservation in Eqs.~(\ref{Hamilton1},\ref{Eq:rescaledE}) in terms of magnetic moments     
\begin{equation}
\frac{1}{2}\; \left({m^{\scriptscriptstyle{(I)}}}^2+ {m^{\scriptscriptstyle{(II)}}}^2\right)= E  \;, \label{Eq:mapping00}
\end{equation}
suggests to  parametrize them by the angle $\varphi$ 
\begin{eqnarray}
m^{(I)}&=&\sqrt{2 E} \cos(\varphi)\cr
m^{\scriptscriptstyle{(II)}}&=&\sqrt{2 E }\sin(\varphi)\;,\label{Eq:mapping0}
\end{eqnarray}
the corresponding state space of which is the unit-circle $\mathscr{S}^1$. This  new non-bounded space $\mathscr{S}^1 $ is mapped onto the bounded interval of energy exchange (Eq.~(\ref{Eq:intervalstatespace})) by,
\begin{eqnarray}
\mathscr{S}^1&\to & \mathscr{I}\cr
\varphi &\to & \Delta E=E (\cos^2(\varphi)-\cos^2(\varphi_0))\;, \label{Eq:mapping}
\end{eqnarray}
with $\varphi_0$ giving the initial magnetic moments  in Eq.~(\ref{Eq:mapping0}) shown in Fig.~\ref{Fig:GFSweakv}(c). 
With Eqs.~(\ref{Eq:maxEntropyGFSGFS2},\ref{Eq:mapping0}) the maximum entropy is located  at $\varphi_{max}=\pm \pi/2$ (Fig.~\ref{Fig:GFSweakv}(d)). Notice, that two opposing magnetic moments, and, hence, angles, map into one state of energy. As the unitcircle is unbound these maxima are local ones (Fig.~\ref{Fig:GFSweakv} (d)), i.e. 
\begin{equation}
\partial_\varphi S|_{\varphi_{max}}=0\;.\label{Eq:Sphi}
\end{equation}  
Though mathematically almost trivial, from our notion of thermodynamics this is at a first glance counter intuitive. 
One could define state variables $\phi^{(i)} $ of system $i$ by
\begin{equation}
\phi^{(i)}=\arctan(m^{(j)}/m^{(i)})_{ j\neq i}\;,\label{Eq:phi}
\end{equation}
with a corresponding temperature
\begin{equation}
\frac{1}{T^{(i)}_{\phi^{(i)}}}= \partial_{\phi^{(i)}} S^{(i)}\;.\label{Eq:phitemp}
 \end{equation}
Likewise energy, this parameter would also have the property of an exchange variable, i.e. $\phi^{(I)} + \Delta \phi$ in system $I$ would imply a change in system $II $ of $\phi^{(II)}-\Delta \phi $.  Notice, that with Eqs.~(\ref{Eq:mapping0}) $\phi^{(I)}=\varphi$ but $\phi^{(II)}=\pi/2-\varphi$.  Insertion into  Eq.~(\ref{Eq:Sphi}) implies an equivalence of these temperatures at the most probable state
\begin{equation}
\frac{1}{T_{\phi_{max}^{(I)}}^{(I)}}=\frac{1}{T_{\phi_{max}^{(II)}}^{(II)}}\;.
\end{equation}
Therefore, one could get the impression that  the choice of macroscopic parameters determines whether the two systems have different or equal temperatures in their most probable state. This apparent paradox is resolved by comparing this new temperature in Eq.~(\ref{Eq:phitemp}) with the Boltzmann temperature, i.e.
\begin{equation}
\frac{1}{T_{\phi^{(i)}}^{(i)}}=\frac{1}{T^{(i)}}\;\partial_{\phi^{(i)}} E^{(i)}\;.
\end{equation}
The maximum of entropy, i.e. the most probable state,  is when $I$ is in one of its ground states, i.e. at its energetic  minimum - ($\phi_{max}^{(I)}=\pm\pi/2,\; E_{min}^{(I)}=0 $),  and system $II$ is at its maximum energy ($\phi_{max}^{(II)}=0,\; E_{max}^{II)}=E_0^{(I)}+E_0^{(II)} $). Here, for the energies of subsystem holds,  $\partial_{\phi^{(i)}} E^{(i)}=0 $. This makes the alternative inverse temperatures become  equivalent by vanishing at the point of maximum entropy, 
 \begin{equation}
1/T_{\phi_{max}^{(i)}}^{(i)}=0\;.
 \end{equation}
Notice, that this is also evident from a mathematical point of view, as  a smooth re-parametrization  $x\to \Delta E(x) \in \mathscr{I}$ transforming a non-local extrema at the boundaries into a local one, must fulfill $\partial_x E^{(i)}|_{x_{max}}=0$. 

However, though the above re-parametrization is formally consistent, there is a drawback from a physical point of view. The parameter $\phi^{(i)}$ dose not characterize solely the system $i$ but also its interaction partner $j$ (Eq.~(\ref{Eq:phi})). Hence, unlike the Boltzmann temperature, the derived temperature cannot be assigned to one system, but always characterizes both partners.

\section{GFS Weakly Coupled to an Ideal Gas}
\label{section:GFS-Gas}

\begin{figure*}
\onecolumngrid
\centering
\includegraphics[width=16cm] {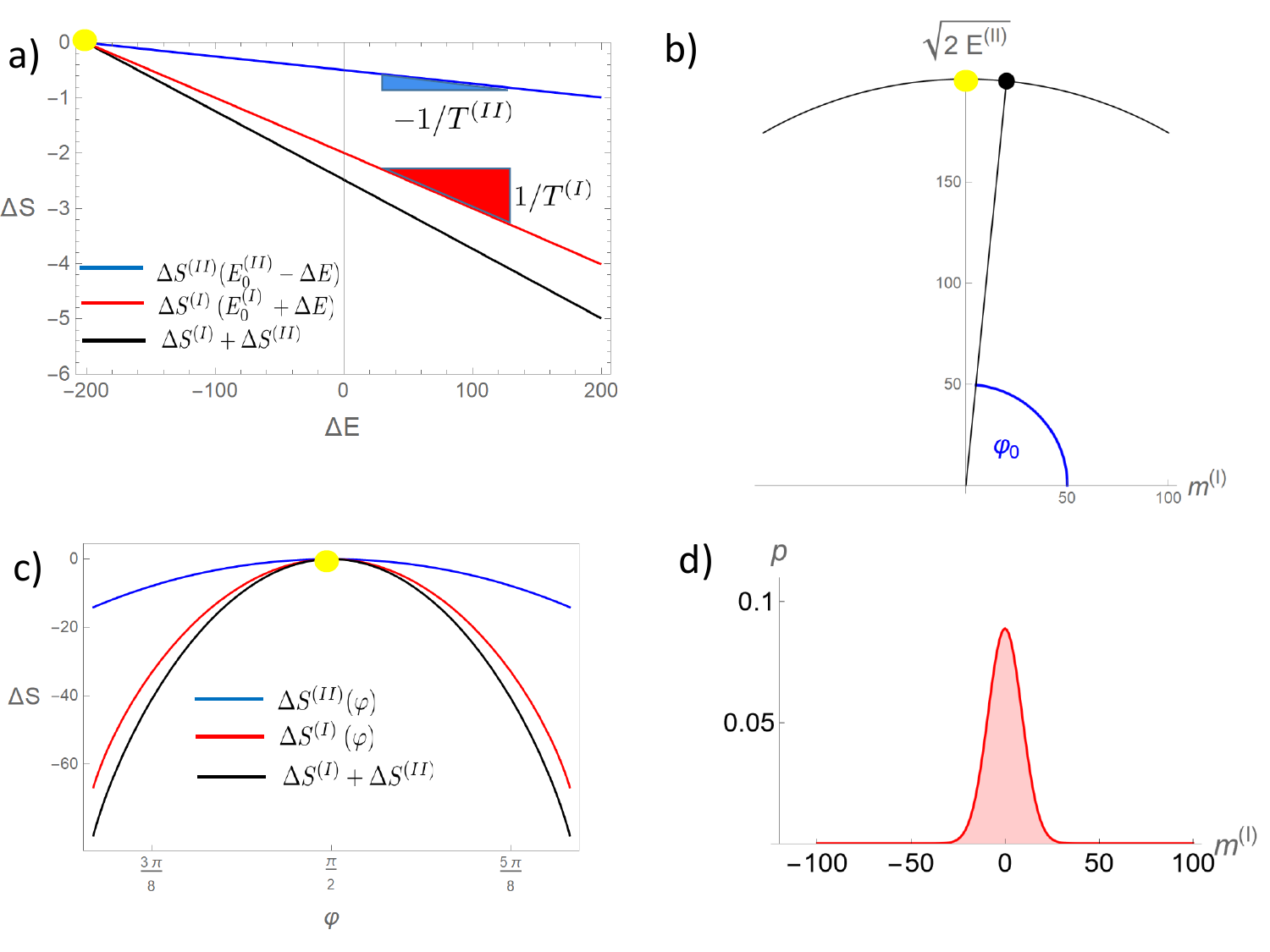}
\caption{A GFS with $ N^{(I)}=100 $ spins is weakly coupled to an ideal 1-D gas consisting of $N^{(II)}=100 $ particles. The initial magnetic moment of the GFS is $m^{(I)}_0=20$, i.e. its  initial energy is $E^{(I)}_0=200 $, that of the gas is chosen to be $  E^{(II)}_0=20 000$.  
a) Normalized entropies of the subsystems and the combined system as a function of the exchanged energy $\Delta E$. Normalization refers to the entropy after delivery of all energy from the GFS to the gas,  $\Delta E=-E^{(I)}_0=-200$, i.e. $S^{(i)}\to \Delta S^{(i)}(E^{(i)})=S^{(i)}(E^{(i)})-S^{(i)}(E^{(i)}_0-E^{(I)}_0)$, with $S^{(i)} $ obtained from Eqs.~(\ref{Eq:BEntropygfsm},\ref{Eq:EntropyGas}).  The absolute maximum of entropy of the combined system is when the GFS is in one of its ground states, $\Delta E=-E^{(I)}_0=-200$ (yellow disk). Here, the temperature of the gas is positive, that of the GFS negative as the respective slope triangles show.  Notice, that the here chosen range of energy exchange, $[-200, 200]$, is sufficiently narrow, so that $\Delta S^{(II)}$ depends approximately linearly on $\Delta E $, and temperature of the gas (Eq.~(\ref{Eq:TempGas})) stays almost constant.  
b)  Reparametrization of energy by  $\varphi \to \Delta E=E (\cos^2(\varphi)-\cos^2(\varphi_0) $, with $\varphi=\arctan(\sqrt{2E^{(II)}}/m^{(I)}) $, and  $E$ as the energy of the combined system. The dark point denotes the initial state $\left(m^{(I)}_0,\sqrt{2E^{(II)}_0}\right) $.  c) Entropies in the new parametrization, with local maximum ($\partial_\varphi \Delta S=0 $) at $\varphi_{max}=\pi/2$.  d) Distribution of probabilities of the magnetic moment, $m^{\scriptscriptstyle{(I)}}=\sqrt{2E}\cos\varphi $, of the GFS in equilibrium. With $\Delta S=\Delta S^{(I)}+\Delta S^{(II)} $ these probabilities hold $p(m^{(I)})\sim \exp(\Delta S(m^{(I))})$.   }
\label{Fig:GFS_Gas1}
\twocolumngrid
\end{figure*}

A GFS which is thermally weakly coupled to a classical ideal gas confronts with the situation, that both partners stay within their temperature regime, the GFS in the negative and the classical gas with the positive one. We will label the gas by the superscript $ II$.  The entropy of a classical 1-dimensional ideal gas,  with $ N^{(II)}$ particles is obtained by the Sackur-Tetrode Equation 
\begin{equation}
S^{(II)}=\frac{N^{(II)}}{2}\; \ln(E^{(II)})+C\;,\label{Eq:EntropyGas}
\end{equation} 
where  $C$ comprises the Planck constant, variables as the volume, which we keep constant, the particle number $N^{(II)}$, and parameters of the $N^{(II)}$-dimensional unit sphere. As they are of no relevance here, the parameter  $C$ is omitted. The well known result for the inverse  Boltzmann temperature is
\begin{equation}
\frac{1}{T^{(II)}}=\partial_{E^{(II)}} S^{(II)}=\frac{N^{(II)}}{2}\frac{1}{E^{(II)}} \;.\label{Eq:TempGas}
\end{equation}     
An increasing  entropy of the GFS and the gas by heat transfer from the first to the latter, makes the drift force in  Eq.~(\ref{driftforce4}) drive the combined system towards the ground states of the GFS, at which it has delivered all its initial  energy $E^{(I)}_0 $ to the gas, $\Delta E=- E^{(I)}_0$ (Fig.~\ref{Fig:GFS_Gas1}(a)). At this  most probable state the temperatures of both subsystems differ. The GFS retains its negative temperature, the gas its  positive. As in the previous section, this global maximum may be transformed into a local one by appropriate re-parametrization. The subsystems' energies $ E^{(I)}, E^{(II)}$  as state parameters for the two systems are replaced by the magnetic moment of the GFS, $m^{(I)} $ and the momentum like quantity $\sqrt{2 E^{(II)}} $. Conservation of energy, $1/2 (m^{(I)})^2+E^{(II))}=E$   implies that the state of combined system can be described by the angel $\varphi$ as 
\begin{eqnarray}
m^{(I)}&=&\sqrt{2 E}\cos(\varphi)\cr 
\sqrt{2 E^{(II)}}&=&\sqrt{2 E}\sin(\varphi)\;,\label{Eq:GFS_Gas_phi}
\end{eqnarray}
as shown in Fig.~\ref{Fig:GFS_Gas1}(b).
The old and new state parameters $\Delta E,\; \varphi $, as well as their respective parameter spaces are related by 
\begin{align}
\tilde{\mathscr{S}}^1&\to  &J &=[-E^{(I)}_0, \min[1/2 (N^{(I)})^2 - E^{(I)}_0, E^{(II)}_0]] \cr
\varphi &\to  &\Delta E &= 1/2 \cos^2(\varphi)-1/2 \cos^2(\varphi_0)
\end{align}
where $\tilde{\mathscr{S}}^1 $ is a segment of the unit circle, the interval $J$ defines the range of energy exchange, and $\varphi_0$ is the angle related to the initial state. Within this new parameter space, entropy has a local maximum $\partial_{\varphi}S=0 $ at $\varphi=\pi/2$ (Fig.~\ref{Fig:GFS_Gas1}(c)).  Similar to Eqs.~(\ref{Eq:phi}, \ref{Eq:phitemp}) in the previous section one may introduce new state parameters $\phi^{(I)},\; \phi^{(II)} $ of the subsystems with associated temperatures $1/T^{(i)}_{\phi^{(i)}}=\partial_{\phi^{(i)}}S^{(i)} $, and $\varphi$ may be interpreted as an exchange parameter. Figure~\ref{Fig:GFS_Gas1}(c) shows that the local maxima of entropies of sub- and  combined system coincide at $\varphi=\pi/2$, i.e.inverse temperatures of the subsystems become equivalent by vanishing, $1/T^{(i)}_{\phi^{(i)}}=0$.  However, the interpretation of the inverse temperature  $1/T^{(i)}_{\phi^{(i)}} $ for system $i$ is problematic, as its conjugated variable $\phi^{(i)}$, in contrast to energy, depends on both subsystems.

\begin{figure}
\includegraphics[width=8cm] {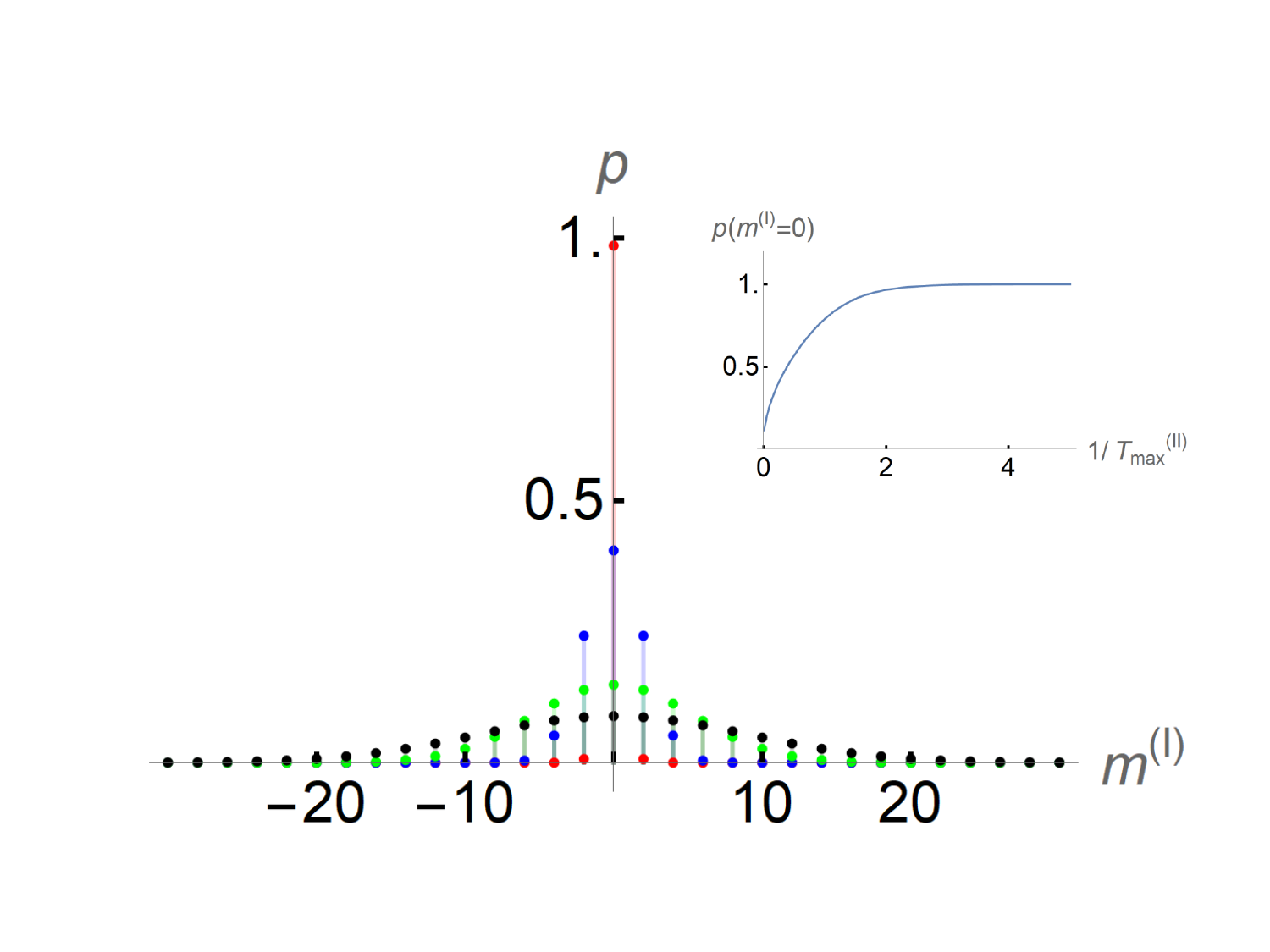}
\caption{Equilibrium distribution of the magnetic moment of a GFS weakly coupled  to an ideal gas. Energies of the GFS and the gas are the same as in Fig.~\ref{Fig:GFS_Gas1}. Particle number of the GFS is held constant at $N^{(I)}=100$, whereas that of the gas, and, hence its temperature, is varied, $N^{(II)}=$100 (black), 1000 (green), 10 000 (blue) and 100 000 (red). The insert shows for $N^{(II)}=100\;000$ particles the equilibrium probability of the GFS to be in one of its ground states, $p(m^{(I)}=0)$ as a function of the temperature of the gas at this point, i.e. with Eq.~(\ref{Eq:TempGas})  $\tfrac{1}{T^{(II)}_{max}}=\tfrac{N^{(II)}}{2} \tfrac{1}{E^{(I)}_0+E^{(II)}_0}$ .} 
\label{Fig:GFS_Gas2}
\end{figure}
The thermodynamics of the combined system of  weakly coupled GFS and classical gas also demonstrates that the most probable state and thermodynamic equilibrium must not be confused.  
The probability distribution of the magnetic moment of the GFS in Fig.~\ref{Fig:GFS_Gas1}(d) shows  a maximum value of only $<0.1 $ at the ground states of the GFS, $m^{(I)}=E^{(I)}=0 $. Therefore there is a good chance to find the combined system in a different energetic state. The more particles the gas at constant energy has, i.e the cooler it is, the more is the probability distribution centered around the ground states of the GFS (Fig.~\ref{Fig:GFS_Gas2}).  With decreasing temperature of the gas,  the probability to find the GFS in its ground states approaches asymptotically one.

\section{GFS Weakly Coupled to a Paramagnetic Polarized Spin Gas}
\label{section:GFSPSGas}

We now couple the GFS (system $I$) to a  gas of $N^{(II)}$ spins, the magnetic dipole moments of which are assumed to be polarized by an external homogeneous magnetic field $B$ (system II). The energy and entropy of this gas with magnetic moment $m^{(II)}$ is
\begin{eqnarray}
E^{(II)}&=&-m^{(II)} B\;, \label{Eq:SpinGasEnergy}\cr
S^{(II)}&=&\log\left(\binom{N^{(II)}}{\frac{N^{(II)}+m^{(II)}}{2}}\right)\;.\label{Eq:SpinGasEntropy}
\end{eqnarray}

\begin{figure*}
\onecolumngrid
\centering
\includegraphics[width=\textwidth] {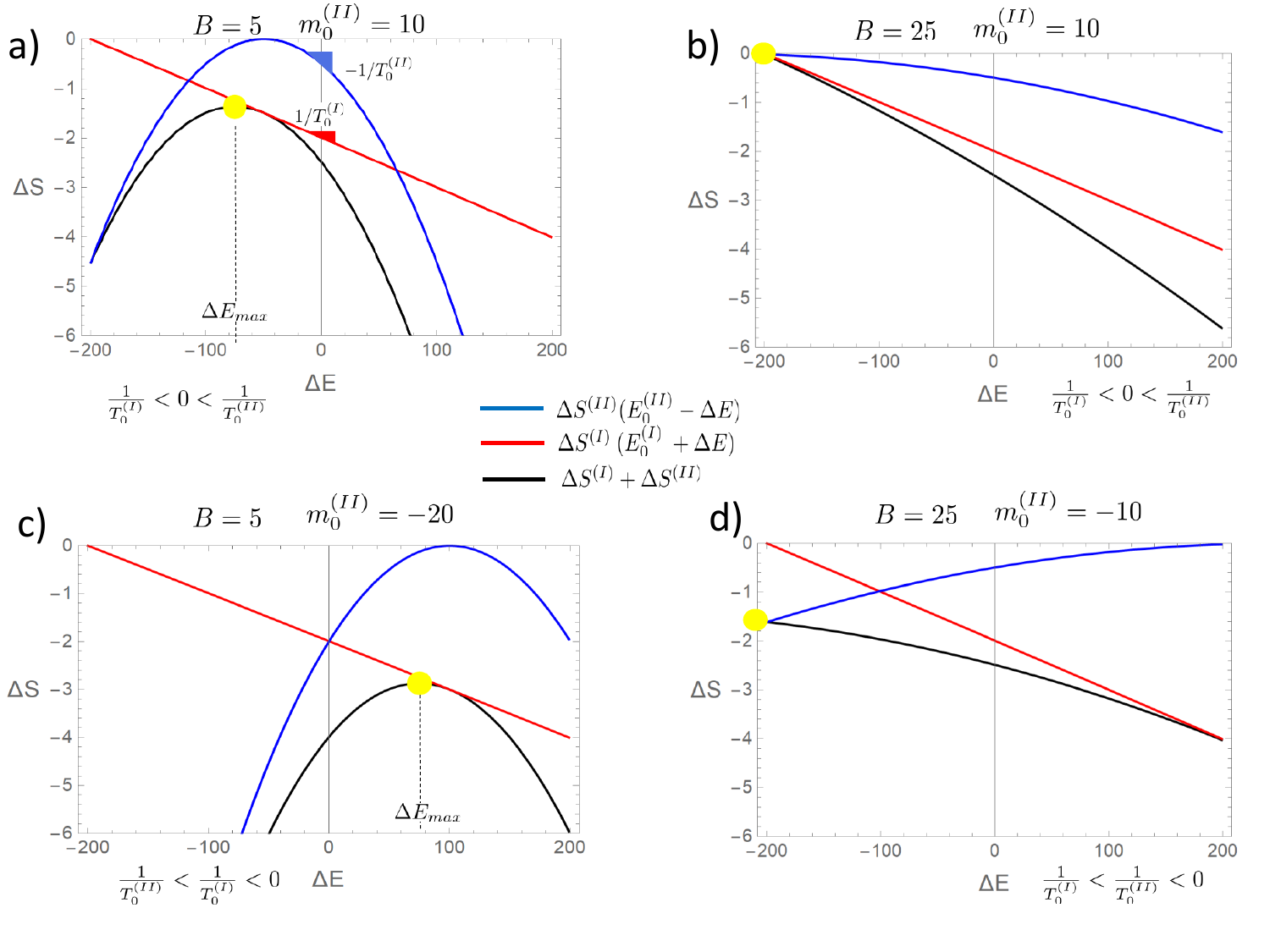}
\caption{A GFS ($N^{(I)}=100 $ spins) in weak thermal contact with a paramagnetic spin gas ($N^{(II)}=100 $ spins). An external magnetic field $B$ polarizes the gas, so that a magnetic moment  $m^{(II)} $ implies an energy $E^{(II)}=-m^{(II)} B$. The entropies of the combined system and subsystems are shown for different temperature scenarios of subsystems  at the beginning ($1/T^{(i)}_0$) of the weak thermal contact, and different values of specific heat of the gas as determined by the magnetic field strengths, $\sim B^2 $ (Eq.~(\ref{Eq:SpinGasTemp})). The initial magnetic moment and hence energy  of the GFS is throughout $m^{(I)}_0=20, \; E^{(I)}_0=200 $, that of the gas $m^{\scriptscriptstyle{(II)}}_0$ is varied and determines its  initial temperature (Eq.~(\ref{Eq:SpinGasTemp})).   a) the initial inverse temperature  of the GFS is smaller than that of the gas (see slope of small triangles), implying heat transfer from the GFS to the gas ($\Delta E <0$) to reach maximum entropy of the combined system. Magnetic field strength is chosen sufficiently low, so that the  small specific heat of the gas enables a local maximum  (see Eq.~(\ref{Eq:localmax})) within the capacity of heat, which  the GFS can transfer ($\Delta E_{max}> -E^{(I)}_0 $). Local maximum of entropy implies that GFS and gas show here equivalent temperatures. b) same as a) but a high $B$ field, and hence magnitude of specific heat of the gas,  implies that a local maximum of entropy would be  beyond the heat that the GFS may transfer, and so  maximum entropy is reached at the ground states of the GFS at which temperatures of both subsystems still differ.  c) The initial magnetic moment of the gas is set anti- parallel to the magnetic field $B$, and sufficiently high in magnitude, so that it becomes the hotter subsystem which transfers heart to the GFS to reach maximum entropy ($\Delta E>0$). Again the $B$ field is low enough so that a local maximum is feasible. d) The initial inverse temperature of the gas is still negative but greater than that of the GFS. A high $B$ field with respective magnitude of specific heat   hampers that the GFS can transfer sufficient heat to the gas to reach a local maximum of entropy. Instead,entropy of the combined system shows an absolute maximum after the GFS has delivered all energy and stays in one of its  ground states. Here, both subsystems exhibit different temperatures.  } 
\label{Fig:SpinGas}
\twocolumngrid
\end{figure*}

Similar to the GFS we define the magnetic moment of the spin gas as the difference of up and down polarized spins.  For a sufficient large number of spins in the gas, the entropy may be approximated by the logarithm of a Gaussian (see Eq.~(\ref{Eq:Gaussian})), i.e. after normalization of the entropy to its maximum $S^{(II)}_{max}$ at $m^{(II)}=0$, 
\begin{eqnarray}
\Delta S^{\scriptscriptstyle{(II)}}&=&S^{\scriptscriptstyle{(II)}}-S^{\scriptscriptstyle{(II)}}_{max} \cr\cr
\Delta S^{\scriptscriptstyle{(II)}}&=&-1/2\; (m^{\scriptscriptstyle{(II)}})^2/N^{\scriptscriptstyle{(II)}} \cr\cr
&=& -1/2 (E^{\scriptscriptstyle{(II)}}/B)^2/N^{\scriptscriptstyle{(II)}}\;,\label{Eq:SpinGasEntropy2}
\end{eqnarray}
and the inverse spin temperature becomes
\begin{equation}
\frac{1}{T^{\scriptscriptstyle{(II)}}}=\frac{\partial\Delta S^{\scriptscriptstyle{(II)}}}{\partial E^{\scriptscriptstyle{(II)}}}=-\frac{E^{\scriptscriptstyle{(II)}}}{ N^{\scriptscriptstyle{(II)}} B^2}\;.\label{Eq:SpinGasTemp}
\end{equation}
In contrast to the previous examples, the temperature of a spin gas covers a range of  positive and negative values, the sign of which changes discordantly with its energy $E^{\scriptscriptstyle{(II)}} $. Hence, the combined system of GFS and spin gas may reach  a most probable state (Eq.~(\ref{driftforce4})) at which both have the same temperature. This most probable state is determined by the local maximum of the combined entropy, (see Eqs.~(\ref{Eq:BEntropygfs}, \ref{Eq:SpinGasEntropy2}) )  
\begin{eqnarray}
\Delta S&=&\Delta S^{(I)}+\Delta S^{\scriptscriptstyle{(II)}} \\
&=&-\frac{E^{(I)}}{N^{(I)}}-\frac{1}{2}\; \frac{(E^{\scriptscriptstyle{(II)}}/B)^2}{N^{\scriptscriptstyle{(II)}} } \;.
\end{eqnarray}
With initial energies $E^{(i)}_0 $ of the respective subsystems  and $\Delta E$ as energy shift of the GFS, this  local maximum( $\partial_{\Delta E}\Delta S=0 $) is located at 
\begin{equation}
\Delta E_{max}=-\frac{N^{\scriptscriptstyle{(II)}}}{N^{(I)}} B^2 +E^{\scriptscriptstyle{(II)}}_0 = E^{\scriptscriptstyle{(II)}}_0 \left(1-\frac{T^{\scriptscriptstyle{(II)}}_0}{T^{(I)}_0}\right)\;,\label{Eq:localmax}
\end{equation} 
with $T^{(i)}_0 $ as the temperatures of the subsystems before thermal contact. This local maximum exists, if $\Delta E_{max} $ lies within  the constraints of energy delivery and uptake between  GFS and spin gas, 
\begin{widetext}
\begin{equation}
-\min(E^{(I)}_0,\;N^{\scriptscriptstyle{(II)}} B-E^{\scriptscriptstyle{(II)}}_0)\le\Delta E_{max}\le 
\min\left(\frac{1}{2} (N^{(I)})^2-E^{(I)}_0, N^{\scriptscriptstyle{(II)}} B+E^{\scriptscriptstyle{(II)}}_0\right)\;.
\end{equation}
\end{widetext}
For simplicity we choose parameters of the spin gas in the following so, that the only constraints for energy delivery of the GFS  is its  initial energy,  $E^{(I)}_0=1/2 (m^{(I)}_0)^2$. 
In Fig.~\ref{Fig:SpinGas}(a) a GFS is brought in contact with  a spin gas which is initially polarized parallel to the magnetic field, i.e its initial energy is negative (Eq.~(\ref{Eq:SpinGasEnergy})) and the corresponding temperature positive (Eq.~(\ref{Eq:SpinGasTemp})). The hotter GFS (negative temperature) delivers heat to the spin gas, $\Delta E<0$, to reach maximum entropy of the combined system. A sufficient small $B$ field, and hence small magnitude of specific heat of the gas ($\sim B^2$, see  Eq.~(\ref{Eq:SpinGasTemp})) guarantees that this state is reached after the gas is  polarized antiparallel to $B$ and has reached the negative temperature of the GFS, i.e. there exists a local maximum of entropy. In contrast, in Fig.~\ref{Fig:SpinGas}(b), a high $B$ field  maintains, by high magnitude of the specific heat, the parallel polarization of the spin gas, i.e. its positive temperature, during energy uptake from the GFS,  until maximum entropy is reached at the latter's ground states, $\Delta E=-E^{(I)}_0$. In Figs.~\ref{Fig:SpinGas}(c,d) the spin gas is polarized anti-parallel to the magnetic field, i.e. it starts with a negative temperature. In Fig.~\ref{Fig:SpinGas}(c), the spin gas is hotter than the GFS ($1/T^{\scriptscriptstyle{(II)}}_0<1/T^{(I)}_0 $), i.e. the gas delivers heat to the GFS,  $\Delta E>0 $, to reach maximum entropy of the combined system.  A sufficient low $B$  field and high capacity of the GFS for energy uptake enables that the inverse temperature of the gas increases (cooling) till it reaches that of the GFS, implying a local maximum. Different, in Fig.~\ref{Fig:SpinGas}(d), a high $B$ field implies that  the hotter GFS ($1/T^{(I)}_0<1/T^{\scriptscriptstyle{(II)}}_0 $) must deliver all its energy until it has reached its ground states, but temperatures have not become equivalent. 

\section{Discussion}
\label{section:Discussion}
\subsection{Boltzmann Entropy, Temperature and Prediction of State}
The discussion about the ``correct'' entropy, and, hence temperature, i.e. Gibbs vs.  Boltzmann version,  is still vivid, through the majority of physicists favors the latter. The advantage of Gibbs (phase space) volume entropy is that based on the adiabatic invariance of phase space volume in statistical mechanics  \cite{rugh2001microthermodynamic,dunkel2014consistent,hilbert2014thermodynamic} it is consistent with fundamental relations in thermodynamics, which, however, this is not a specific feature of Gibbs entropy as there exists some gauge freedom. Any smooth bijejctive map of this entropy, $S_G\xrightarrow{f}\tilde{S} $ with rescaling of temperature $\tilde{T}=T_G/f^{\prime} $ maintains this consistency \cite{campisi2015construction, campisi2016erratum}, in particular, if this holds for  mapping between phase space volume (Gibbs) and its surface (Boltzmann). This, for example, is the case  for a paramagnetic spin systems in an external  magnetic field  \cite{abraham2017physics}, separately for its  positive and negative energy domain. Whether Gibbs entropy is applicable to particle exchange and the chemical potential is also matter of debate \cite{tavassoli2015microcanonical}, which is most probably related to the fact that  Gibbs entropy and derived quantities treat energy and particles in a non-symmetric way.  Gibbs entropy as a function of energy  and particle number considers all  states at this number, but energy may be below or equal. A symmetric approach would also include states with particle numbers below. This approach is motivated as there should be no formal reason, why to treat energy and particle exchange in a different way, which for example becomes best evident from photons, for which energy and particle exchange are directly coupled.   

Gibbs as well as Boltzmann entropy  may, perhaps to different degrees, prove to be consistent with important aspects of statistical mechanics, however, the main  advantage of Boltzmann entropy and temperature is that it predicts the probability of the state of a composite system, the subsystems of which are   weakly interacting. For example, the canonical distribution with predictions of fluctuations, follows directly from Boltzmann's definition of entropy and temperature \cite{cerino2015consistent, buonsante2016dispute}, and this is independent from the sign of temperature.  In particular Boltzmann temperature is equivalent with the temperature in the canonical ensemble, which is not a trivia, as both are  defined differently. In particular, the latter is introduced as a Lagrangian multiplier, to find the most probable energy distribution under the constraint of energy conservation \cite{buonsante2016dispute}. 

Furthermore, equivalence of Boltzmann  temperatures of two systems in weak thermal contact predicts the most probable state, which, for ``normal'' systems, is  equivalent with the equilibrium state,  i.e in accordance with phenomenological thermodynamics. 

\subsection{Temperature Gradient as Drift Force for Weakly Coupled Subsystems}
In this paper we applied Boltzmann's approach within the framework of statistical thermodynamics. Time reversal symmetry of dynamics in equilibrium implies detailed balance, and this principle makes explicit, how gradients of Boltzmann entropies become  stochastic driving forces. For a system, which is composed of thermally weakly coupled subsystems, these entropy gradients translate into temperature gradients of the subsystems as driving forces. With this general tool in hand, one is able to tackle also questions the answers of  which are at a first glance outside our normal notion of thermodynamics; e.g. how do systems behave, for which temperature is  independent from its energetic state, or whether and how is equilibrium established between two systems which exist in different temperature domains, e.g. at positive and negative values.  Closely related are implications for temperature measurements of - and with systems, which exist only at negative temperatures, and accordingly, the question, whether the zeroth law of thermodynamics is still valid here?      

It must be stressed that for the above derivation no presupposition of the weakly coupled subsystems were made. In  this manuscript a special class of strongly interacting spin systems, the GFSs, were considered. These can be thought of as $N$ interacting spins positioned at the vertexes of a regular $(N-1)-$dimensional simplex, or as spins of a lattice with very long ranging interactions. Related to the latter notion, a similar interaction is frequently applied for modeling ferromagnetism by the Curie-Weiss Hamiltonian \cite{campa2009statistical}. However, in contrast to the here considered GFS Hamiltonian $H$ in   Eq.~(\ref{Hamilton1}), the Curie-Weiss Hamiltonian is scaled by the number of spins, i.e. $H_{CW}=H/N $, making energy and entropy extensive -, though still not additive, quantities, and temperature an intensive parameter.  However, such a re-scaling would have been an unnecessary  constraint. In fact, as we only consider thermal coupling between subsystems, the statements in this manuscript would also hold for such a re-scaled, antiferromagnetic  Curie-Weiss Hamiltonian. Similar to a GFS, temperature would be negative and independent from energy,  but  independent from the number of spins. Nevertheless different temperatures could be achieved by different coupling constants $J$. 

One might argue, that in this manuscript thermodynamics of strongly interacting systems (GFS) is mixed with conventional thermodynamics of weakly coupled systems. However, this is not correct. In fact we do not consider thermodynamics in the sense of energy exchange and temperature distribution within a strongly interacting system. Here, of course, traditional measures as Boltzmann entropy and temperature have its limitation, and other entropies come into play. Instead, we consider such a system from outside having macroscopic state parameters (energy, magnetization) which derive from the corresponding microstates, the number of which is quantified by the Boltzmann entropy. The only presupposition for these systems  is, that in thermodynamic equilibrium time reversal symmetry for stochastic transitions between microstates holds, and that the latter have  equal a priori probability.  These systems are brought in weak thermal contact with another system; another GFS, an ideal gas or a paramagnetic spin gas. The weak coupling ensures additivity of entropy, with respect to the macroscopic parameters, e.g. energy, of the subsystems.  Thermal weak coupling of strongly interacting subsystems, as we did,  is nothing new, but frequently applied in various fields,  e.g. to demonstrate ensemble inequivalence by coupling to a bath \cite{baldovin2018physical} or for showing the limitation of the zeroth law of thermodynamics by weak coupling of two systems with negative specific heat \cite{ramirez2008systems}. 

\subsection{The Zeroth Law of Thermodynamics}
\label{subsection:zeroth}
The examples in which a GFS was coupled to another GFS or an  ideal gas showed that equilibrium of the combined systems may be present with the subsystems being at different temperatures. This demands reconsideration of the zeroth law of thermodynamics.

The zeroth law of thermodynamics states that if two  systems are in thermodynamic equilibrium with a third, all systems are in thermodynamic equilibrium with each other, which makes ``being in equilibrium'' a transitive relation. In terms of phenomenological thermodynamics equilibrium implies constancy of energy of the subsystems, and hence,  absence of heat transfer. The system is stable against small exchanges of energy between subsystems, making this a reversible process. Hence entropy of the combined system is stationary and the subsystems' intensive parameter `` temperature '' becomes equivalent.  This is the base for temperature measurements by a thermometer. Two isolated subsystems, the temperatures of which were measured to be equivalent, will not exchange  heat, after having been brought in contact. Therefore their energies will remain constant at their initial value, i.e. they are in equilibrium.  

To understand the implications for the GFSs, a detailed reconsideration of the above, apparently trivial, concepts and statements in terms of statistical mechanics is required. From an ensemble view, a system in equilibrium exhibits equal probability of all its microstates $\sigma$. If these microstates translate into macroscopic variables, i.e. in our case energies of the  two subsystems in weak  contact, $\sigma\to (E^{(I)},E^{(II)})$, this implies that the probability to find the state $(E^{(I)},E^{(II)}) $ is proportional to the number of microstates fulfilling the energetic constraint, i.e. $\sim |\bSigma_{(E^{(I)},E^{(II)})}|=e^{S (E^{(I)},E^{(II)})} $. 

The weak contact assures, that the probability of the state $(E^{(I)},\; E^{(II)})$ is then proportional to the product of its number of microstates at respective energies of the subsystems, i.e.  $\sim \exp(S^{(I)}(E^{(I)})+S^{\scriptscriptstyle{(II)}}(E^{(II)})) $. 
For most systems with many degrees of freedom this product, under the constraint of energy conservation,  peaks narrowly around a most probable state $(E^{(I)},\; E^{(II)})_{max} $, i.e. the combined system stays almost only  in this state. Of course there are fluctuations around this most probable state, however, the strong drift force (Eq.~(\ref{driftforce4})) arising from the steep gradient of entropy near its peak, keeps this fluctuations small. Detailed balance then makes any net energy exchange vanish. As the entropy of the combined system almost stays solely at its local maximum with equal temperatures of subsystems, this  equivalence will be retained after separation.  
If one of these systems, now labeled subsystem 1,  serves as  thermometer, our notion is, that it has measured temperature of subsystem 2. Now, if the thermometer is brought in contact with a 3rd  subsystem, which is assumed to have the same temperature as the thermometer, this combined  state is already  in the  most probable state of the new composite system. Being in its most probable state at the beginning of contact, this composite system will in fact also relax to equilibrium. However, almost all of the equilibrium micro states are also elements of the most probable macro state, i.e. the effect of relaxation is negligible and the composite system appears static from beginning of contact. One states, that subsystem 2 were,  and subsystem 3 is in equilibrium with the thermometer,  the state of which, remained (almost) constant throughout the process. As subsystem 2 and 3  have the  same temperature, their combined state after brought in contact  will be the most probable one. Again relaxation to equilibrium implies only  negligible probability of states beside the most probable one, i.e. system 2 and 3 appear from beginning of contact constant in time, leading to the statement, that they are in equilibrium, and the zeroth law holds.  

The concept of the above process is mainly based on two assumptions,  normally taken for granted. Two subsystems being in weak contact and thermal equilibrium almost solely stay in the most probable state of the combined system, i.e the term`` equilibrium'' and ``most probable state" are used synonymously. And, the second assumption, this most probable state is a local maximum of entropy (or probability) related to  energy exchange between subsystem. The first assumption assures, that after separation of the two subsystems, their respective energies are those, they have in the most probable state. The second assumptions guarantees that the subsystems exhibit equal temperatures in the most probable state which they retain after separation.   

The GFSs challenge this concept, as neither the terms ``most probable state''  and equilibrium are interchangeable nor does the combination of a GFS  with another systems always have  a local maximum of entropy in its most probable state. We will consider the case of a GFS in contact with an ideal gas A. 
Here, the most probable energetic state of the combined system is given, when  the GFS has delivered all its energy and  is in one of its ground states. This holds for contact with any gases at each temperature. However, this most probable state does not  necessarily dominate much the other states. Figure~\ref{Fig:GFS_Gas2} shows that for a sufficient hot gas being in equilibrium with a GFS there is a high chance to find the GFS not in one of its ground states. 
Let us assume that the GFS is taken as a thermometer for ideal gases, i.e. the energy of the gases  are sufficiently high so that a contact with a GFS does not change the gases' temperatures, which is always a reasonable assumption for any thermometer. 
After the thermometer is separated from gas A, with which it had been in thermodynamic equilibrium, it has its negative temperature (Eq.~(\ref{TemperatureGFS})), independent from its energetic state. 
Brought in contact with another gas B of the same temperature as the first gas A, our normal notion (see above) is that the combined system (GFS, gas B)  should be in equilibrium. 
However, such a statement does not make sense, as this initial state of the  GFS and gas B is only one, among others with non-negligibly probability in  equilibrium to which
the newly combined system will  relax. Or loosely speaking: in this example relaxation to equilibrium will most probably change the state of the  thermometer.   Hence, after separated from gas B, the GFS thermometer will probably be in another energetic state than it was after separation from gas A and before  contact with gas B. Hence, there  is potentially energy transferred between the gases, which is definitely not in the sense of the zeroth law, with respect to transitivity of equilibrium. 
However, the concept of the zeroth law makes sense, if we also consider the process of equilibration of thermometer and gases    
as a statistical process. 
The probability to find the GFS in some state with energy $E$ after being separated from a gas, with which it was in equilibrium  is 
\begin{eqnarray}
p(E)&=&C\;\exp\bigg(S^{\scriptscriptstyle{(GFS)}}(E)+S^{\scriptscriptstyle{(gas)}}(E^{(\scriptscriptstyle{(gas)}}-E)\bigg)\; ,\cr\cr
&=&\tilde{C}\;\exp\bigg(\left(\frac{1}{T^{\scriptscriptstyle{(GFS)}}}-\frac{1}{T^{\scriptscriptstyle{(gas)}}}\right) E\bigg)\;, \label{Eq:GFS_Gas_Eq} 
\end{eqnarray}
with $E^{(gas)}$ as energy of the gas, when the GFS is in one of its ground states, and $C,\;\tilde{C}$ as normalization constants. Notice, that $\tilde{C} $ solely depends on temperatures of the GFS and gas. 
We applied in Eq.~(\ref{Eq:GFS_Gas_Eq}) that the GFS working as a thermometer does not affect the temperature of the gas within the  range of energy exchanged near $E^{(gas)} $, i.e $T^{(gas)}$ is independent from $E$. 
Hence, repetitive contacts with equilibration  and separation of thermometer (GFS) and gas reveal the above probability distribution of energy of the GFS. Notice that this probability distribution could also be obtained from a single measurement of the gas by an ensemble of GFSs.  Knowing the temperature of the GFS,  the temperature of the gas may then directly be obtained from the probability distribution. This  procedure, when first applied  for gas A and thereafter for B allows determination of temperature of each. Obviously, 
if the  gases  A and B are  measured to have  the same temperature, and are thereafter brought in contact, there will be no net energy exchange, making the the zeroth law of thermodynamics valid.

To tackle  the problem of potential energy transfer by the thermometer it is worth to consider an alternative procedure  of temperature measurement. 
Alternately equilibration of the   GFS  with gas A   and B,  and repeating this cycle many times will  also reveal the temperature of each gas from the energy spectra of the GFS. 
We consider the energy transfer from A to B in one cycle. After equilibration with gas A the GFS exhibits energy $E^{(A)}$, and thereafter with gas B the energy  $E^{(B)} $. Hence the energy transfer  is
\begin{equation}
\Delta E_{A\to B}=E^{(A)}-E^{(B)}
\end{equation} 
The average energy the thermometer transfers from gas A to B  is the obtained with Eq.~(\ref{Eq:GFS_Gas_Eq})  as 
\begin{widetext}
\begin{eqnarray}
\overline{\Delta E}_{A\to B}&=&\int_0^{E_{m}} (p_A(E)-p_B(E))\; E\; dE \cr\cr
&=&\Delta\beta_B^{-1}-\Delta\beta_A^{-1}+
\frac{E_{m}}{2}\left(\exp\big({\scriptstyle{\frac{E_{m}\Delta\beta_A}{2}}}\big)\sech\big({\scriptstyle{\frac{E_{m}\Delta\beta_A}{2}}}\big)-\exp\big({\scriptstyle{\frac{E_{m}\Delta\beta_B}{2}}}\big)\sech\big({\scriptstyle{\frac{E_{m}\Delta\beta_B}{2}}}\big)  \right)\label{Eq:EX}\;,
\end{eqnarray}
\end{widetext}
with 
\begin{equation}
\Delta\beta_{i}=\frac{1}{T^{\scriptscriptstyle{(GFS)}}}-\frac{1}{T_{i}^{\scriptscriptstyle{(gas)}}}\;,\;\; i=A\; B\;,
\end{equation}
and $E_{m}$ as upper boundary of energy of the GFS thermometer (Eq.~(\ref{Eq:rescaledE})).
Consistently the sign of energy exchange  is parallel with the temperature gradient of the gases, i.e. for equal temperatures there is no exchange.  

The above considerations show that the zeroth law of thermodynamics is also valid for the GFS, however it emerges as a statistical phenomena in repetitive measurements.

\subsection{Measuring Temperature of  a GFS by a Paramagnetic Gas}
As the GFSs have constant negative temperatures its measurement in the classical sense should require a thermometer capable to be in states of negative temperatures as well, e.g a  polarized paramagnetic spin gas, the temperature domain of which ranges from positive to negative values.  As shown in Figs.~\ref{Fig:SpinGas}(a,c)   
a weakly coupled  GFS and spin gas, with initial energies $E_0^{(I)},\; E_0^{(II)} $ , may, after the exchange of energy $\Delta E_{max} $, i.e  $E_0^{(I)}\to E_0^{(I)}+\Delta E_{max},\; E_0^{(II)}\to E_0^{(II)}-\Delta E_{max}$,  exhibit a local maximum of entropy, and, hence, a most probable state with equal temperatures, provided that the  energy range of the GFS and the gas sufficiently match. 

What is important is the fact that this amount of energy exchange $\Delta E_{max}$  does not depend on the initial  energy of the GFS, but only on its temperature $T_0^{(I)}$  (Eq.~(\ref{Eq:localmax})). 
Hence, the combined system of a spin gas (thermometer) and a GFS at the same temperature will always be in its most probable state, independent from the energetic state of the GFS. This has implications for the spin gas, when used as a thermometer in the usual way, which determines temperature from the most probables state (see Subsection \ref{subsection:zeroth}).  For example: if two GFSs at same temperature but at different energetic states are brought in contact with a spin gas thermometer being at the same temperature as the GFSs, each combined system (GFS and thermometer) is already in the most probable state, i.e. independent from the energy of the respective GFS.  At a first glance this may appear puzzling. However, in the concept of inverse temperature differences becoming  drift forces in Eq.~(\ref{driftforce4})  this is consistent. It just reflects the transitivity of ``being  in the most probable state'', as long as this most probable state derives from a local maximum of entropy.

\subsection{Systems at Negative Temperature - Implications for Stability }
In this work we focused on special GFSs as paradigm because its entropy obtained from the number 
of microstates decreases monotonically  at constant slope with its energy, i.e. it exhibits a constant -,  and negative temperature, which are two unfamiliar properties. The property to have solely negative temperatures makes GFS  have features different than the paramagnetic spin gas, which serves normally as the paradigm for negative temperatures. Being at a  negative temperature the latter will, after brought in contact with a ``normal'' system at positive temperature and an unbounded upwards energy
 spectrum, always start to release energy to the ``normal'' system. The spin system will pass the point of its maximum entropy, where its inverse temperature vanishes, and continue to transfer energy 
 till both system have the same positive temperature. This has lead to the statement that equilibrium is never possible at  negative temperatures  \cite{romero2013nonexistence}. A GFS, in contrast to a spin gas,  has solely negative temperatures. Brought in contact with a ``normal'' system,  the different signs of (inverse) temperatures synergistically make  the force in Eq.~(\ref{driftforce4}) drive the combined system to the ground states of the GFS.  As the combined systems has relaxed, i.e probability of  states has become  $\sim e^{S}$ (Fig.~\ref{Fig:GFS_Gas2}), both systems are in thermodynamic equilibrium, though they exhibit different temperatures. Hence, a system at negative temperatures can be in thermodynamic equilibrium with a system at positive temperature.

\section{Conclusion} 
We showed that in weakly thermal  coupled systems gradients of respective inverse Boltzmann temperatures act as drift forces, which, by exchange of energy, drive the combined system towards a most probable state where entropy is at its maximum. Applied to a GFS in contact with another GFS of different temperature or an ideal gas  this most probable state is an absolute but not a local extremum, at which the hotter GFS is in one of its ground states. 
At this most probable state both systems maintain different temperatures.
This implies that our normal notion of equal temperatures at this point of maximum entropy must be reconsidered, which  requires a reformulation of the zeroth law of thermodynamics.  A GFS, when used as a thermometer, does not provide the temperature of the measured system from a single measurement. Instead the probability distribution of energy of the GFS takes its place. This can be obtained by repetitive measurements by a single GFS, or by a single measurement with an ensemble of GFSs. The probability distribution as  ''thermometer parameter ´´ fulfills in particular the requirement of transitivity, i.e.  the zeroth law of thermodynamic becomes valid again. 

The non-local maximum of entropy, with emergence of different temperatures of subsystems at this point, is not a unique feature of the GFSs. It occurs, when the temperature ranges of the subsystems do not match with the energetic constraints to  achieve a local maximum entropy by heat transfer. In particular this always holds for  systems, which stay in different temperature domains, e.g. in our case for negative (GFS) -, and positive (gas) values, but also when energy transfer is constrained as shown for some scenarios of the GFS in contact with a paramagnetic spin gas. 
 
A paramagnetic spin gas  may potentially serve as a thermometer for a GFS, as its temperature range  covers positive and negative values, and, hence,  local extrema of entropy, with equivalent (negative) temperatures become feasible.  
Hence, measurement of the temperature of a GFS by a spin gas is feasible in the usual way. However, such a measurement would not  provide any information about the energetic state of the GFS.  

\appendix
\section{Entropy Difference as Drift force}
\label{appendix1}
The systems we have in focus are isolated and completely described by a microscopic state variable $\bsigma $, which in our case is the spin configuration. The system undergoes stochastic transitions, which form a trajectory $\bsigma(t)$ in state space $\Omega$. For simplicity, it is assumed that it contains a finite number of microstates. Each microstate is accessible from any other by some trajectory (irreducibility), and ergodicity holds. \footnote{Description of systems by means of stochastic thermodynamics mainly refers to objects  the dynamics of which results from thermal contact with some bath \cite{seifert2012,broeck2013,Bauer2013,bauer2017,bauer2020}. However, this approach may also describe complex dynamics of isolated systems, e.g. in order to evaluate the microcanonical ensemble \cite{bhanot1984,ray1991,ray1996,Oliveira2015,Oliveira2019}.}
We now consider an ensemble of such isolated systems, which enables to apply tools from statistics. The $n-$point joint probability distribution function $p(\bsigma_n(t_n),.., \bsigma_1(t_1)) $,    
is then the probability of the path -, or more precisely for a continuous time, the probability of the class of paths with supporting points   $\bsigma_n(t_n)..\bsigma_1(t_1) $. For simplicity we will still refer to this sequence in state space as path or trajectory.  
The dynamics is assumed to be stationary, i.e. these probabilities remain constant under a shift in time $t_i\to t_i+\Delta t,\; i=1\hdots n$. In equilibrium the arrow of time is absent, i.e. time reversal symmetry holds for the joint probabilities,
\begin{equation}
p^{(eq)}(\bsigma_n(t_n),\hdots, \bsigma_1(t_1))=p^{(eq)}(\bsigma_1(t_n),\hdots, \bsigma_n(t_1))\;.
\end{equation}
Together with the assumption of a stationary transition dynamics this implies that the probabilities of a specific trajectory and its time reversal counterpart are equal, i.e. micro reversibility holds. Hence, flow of any physical quantity related to transitions in state space, e.g. flow of energy, vanishes. Applying this time reversal symmetry to the 2-point joint probabilities directly reveals detailed balance as, 
\begin{equation}
p^{(eq)}(\bsigma_2(t_2), \bsigma_1(t_1))=p^{(eq)}(\bsigma_1(t_2), \bsigma_2(t_1))\;. \label{Appendix:2ptimesymmetry}
\end{equation}
This detailed balance is not restricted to transition between single states $\bsigma$, but it also holds for transitions between assemblies of states $\bSigma=\{\bsigma_i\} $, i.e. subsets of $\Omega $. Here, the  2-point joint probability is the probability to find the system at $t_1$and  $ t_2 $  in microstates of the corresponding assemblies. It is therefore obtained by summing up over the joint probabilities of respective microstates  (marginalization)
\begin{equation}
p^{(eq)}(\bSigma_2(t_2), \bSigma_1(t_1))=\sum_{\substack{\bsigma_2\in\bSigma_2(t_2) \\ \bsigma_1\in\bSigma_1(t_1)}} p^{(eq)}(\bsigma_2(t_2), \bsigma_1(t_1))\;,\label{Appendix:2passembledstates}
\end{equation} 
and with Eq.~(\ref{Appendix:2ptimesymmetry}) and one gets
\begin{equation}
p^{(eq)}(\bSigma_2(t_2), \bSigma_1(t_1))=p^{(eq)}(\bSigma_1(t_2), \bSigma_2(t_1))\;. \label{Appendix:2ptimesymmetryassembledstates}
\end{equation}
With the conditional probability $p(\bSigma_i|\bSigma_j;\Delta t)$ for a transition $\bSigma_i\leftarrow\bSigma_j$, i.e. to find the system after $\Delta t=t_2-t_1$ in a microstate of $\bSigma_i $ if it was at  $t_1$ in a microstate of  $\bSigma_j $, and the equilibrium probability $p^{(eq)}(\bSigma)=\sum_{\bsigma\in\bSigma} p^{(eq)}(\bsigma)$ of the assembly $\bSigma$ one obtains   
\begin{eqnarray}
p^{(eq)}(\bSigma_2(t_2), \bSigma_1(t_1))&=&p(\bSigma_2|\bSigma_1;\Delta t)p^{(eq)}(\bSigma_1) \cr\cr
&=&p^{(eq)}(\bSigma_1(t_2), \bSigma_2(t_1))\cr\cr
&=&p(\bSigma_1|\bSigma_2;\Delta t)p^{(eq)}(\bSigma_2) \;.
\label{Appendix:Eq:Fundamental}
\end{eqnarray}  
 The latter Equation may be rewritten as
 \begin{equation}
 \underbrace{\frac{p(\bSigma_2|\bSigma_1;\Delta t)} {p(\bSigma_1|\bSigma_2;\Delta t)}}_{\text{probalistic drift dynamics}}=\underbrace{\frac{p^{(eq)}(\bSigma_2)}{p^{(eq)}(\bSigma_1)}}_{\text{ equilibrium parameters}}\;.\label{Appendix:ForceEntropy}
 \end{equation}
This Equation connects drift dynamics of the system on the left hand side, to static measures of equilibrium on the right hand side.  On the left are the conditional probabilities which quantify the transitions between the two assemblies  within a time interval $\Delta t$, i.e. both  characterize the dynamics of the system. Notice, that by this relation the ratio of conditional probabilities becomes independent from time.

\section{Drift in Space of  Macroscopic State Variables of Subsystems}
\label{appendix2}
We consider the case that the system is composed of subsystems, which are characterized by macroscopic state variables, e.g. volume, particle number magnetization or energy. Subsets of microstates  $\bSigma$  may then be related to macroscopic state observables of the latter, denoted by $\epsilon=(\epsilon_1\cdots \epsilon_n)^t $, with $\epsilon_i$ as the macroscopic state variable of subsystem $i$. As all microstates translate uniquely into a macroscopic variable $\bsigma\to\epsilon(\bsigma) $  one gets   
\begin{eqnarray}
\bSigma_\epsilon\cap\bSigma_{\epsilon'}&=&\emptyset\; \text{for}\; {\epsilon \neq \epsilon'}\;, \label{Appendix:disjoint} \cr\cr
\Omega &=&\bigcup\limits_{\epsilon|_{{\text{\tiny{constraints}}}}} \bSigma_\epsilon\;,   \label{Appendix:complete}    
\end{eqnarray}
i.e. state space $\Omega$ is composed of disjoint subsets characterized  by  macroscopic state parameters of its subsystems.  
The index `$|_{\text{ constraints}}$' indicates that the macroscopic parameter $\epsilon$ may underlie certain constraints, e.g. conservation of energy or particle number, e.g. $\sum\epsilon_i=\text{const}$.
The drift force  in Eq.~(\ref{driftforce}) between two infinitesimal neighbored states $\epsilon '=\epsilon+d\epsilon\leftarrow \epsilon$, takes the differential form  $\mathfrak{F}\to \delta\mathfrak{F}$, as    
 \begin{equation}
\delta \mathfrak{F}_{\epsilon',\epsilon} =dS=\nabla_\epsilon S(\epsilon)\; d\epsilon|_{\text{constraints}} \;,\label{Appendix:driftforce}
\end{equation}     
with $\nabla_\epsilon=(\partial_{\epsilon_1},\cdots,\partial_{\epsilon_n})$.
This makes explicit, that the drift force derives from a potential, namely the entropy, which makes it a conservative force, i.e. $\delta\mathfrak{F}=d\mathfrak{F}$ becomes a total differential.

\section{Factorization of State Space for Weakly Coupled Subsystems}
\label{appendix3}
In energetically weakly coupled subsystems, the interaction is assumed to be negligible with respect  to affect the dynamics between microstates within the respective subsystems.  Therefore subsystems may be treated as if isolated, and the combined system is completely characterized by its energy distribution of subsystems $\epsilon=(E^{(1)},\cdots E^{(n)})^t $.  The subset of microstates for this distribution  $\bSigma_\epsilon$ then factorizes into subsets of the individual subsystems of the respective energies $E^{(i)} $, and may be written as tensor product
\begin{equation}
\bSigma_\epsilon= \bigotimes_{i=1}^{n} \bSigma^{(i)}_{E^{(i)}}\;. \label{Appendix:ProductSpace}
\end{equation}
Therefore the Boltzmann like entropy of the composite system becomes the sum of the Boltzmann entropies of the subsystems
\begin{eqnarray}
S(\bSigma_\epsilon)&=&\sum_{i=1}^{n}S^{(i)}\left(\bSigma^{(i)}_{E^{(i)}}\right)\;, \text{ or simply}\\
&=&\sum_{i=1}^{n}S^{(i)}(E^{(i)}) \label{Appendix:EntropySub}\;.
\end{eqnarray}

\section*{Acknowledgement}
The authors gratefully acknowledges stimulating and fruitful discussions with Marco Baldovin and Andrea Puglisi.  

 \bibliographystyle{apsrev4-2}
\bibliography{literature}

\end{document}